%% file: main.tex
\documentclass[sigconf]{acmart} 

\usepackage{colortbl} 
\usepackage{listings} 

\lstdefinestyle{koliPrompt}{
  basicstyle=\ttfamily\small,breaklines=true,breakatwhitespace=true,breakindent=0pt,breakautoindent=false,columns=fullflexible,keepspaces=true,showstringspaces=false
}

\copyrightyear{2026}
\acmYear{2026}
\setcopyright{cc}
\setcctype{by-nc-nd}
\acmConference[Koli Calling '26]{26th Koli Calling International Conference on Computing Education Research}{November 05--08, 2026}{Koli, Finland}
\acmBooktitle{26th Koli Calling International Conference on Computing Education Research (Koli Calling '26), November 05--08, 2026, Koli, Finland}
\acmDOI{10.1145/3856208.3856226}
\acmISBN{979-8-4007-2752-8/2026/11}

\begin{document}

\title[Correctness, Convergence, and AI-Generated Code Detection: A Longitudinal Study]{Correctness, Convergence, and AI-Generated Code Detection: A Longitudinal Study of Student and Large Language Model Code in Introductory Programming}

\author{Runlong Ye}
\orcid{0000-0003-1064-2333}
\affiliation{%
  \institution{University of Toronto}
  \city{Toronto}
  \state{Ontario}
  \country{Canada}}
\email{harryye@cs.toronto.edu}

\author{Jing Fan}
\orcid{0009-0002-1170-6693}
\affiliation{%
  \institution{Aalto University}
  \city{Espoo}
  \country{Finland}}
\email{jing.fan@aalto.fi}

\author{Angela Zavaleta Bernuy}
\orcid{0000-0002-1228-5774}
\affiliation{%
  \institution{McMaster University}
  \city{Hamilton}
  \state{Ontario}
  \country{Canada}}
\email{zavaleta@mcmaster.ca}

\author{Oscar Karnalim}
\orcid{0000-0003-4930-6249}
\affiliation{%
  \institution{Maranatha Christian University}
  \city{Bandung}
  \state{West Java}
  \country{Indonesia}}
\email{oscar.karnalim@it.maranatha.edu}

\author{Paul Denny}
\orcid{0000-0002-5150-9806}
\affiliation{%
  \institution{University of Auckland}
  \city{Auckland}
  \country{New Zealand}}
\email{p.denny@auckland.ac.nz}

\author{Juho Leinonen}
\authornote{Both authors contributed equally as senior authors.}
\orcid{0000-0001-6829-9449}
\affiliation{%
  \institution{Aalto University}
  \city{Espoo}
  \country{Finland}}
\email{juho.2.leinonen@aalto.fi}

\author{Michael Liut}
\authornotemark[1]
\orcid{0000-0003-2965-5302}
\affiliation{%
  \institution{University of Toronto Mississauga}
  \city{Mississauga}
  \state{Ontario}
  \country{Canada}}
\email{michael.liut@utoronto.ca}

\renewcommand{\shortauthors}{Ye et al.}

\begin{abstract}
Large language models can generate plausible solutions to programming assignments, making it tempting to detect their use by matching student code against a reference bank of generated solutions. Yet similar code can also arise when an assignment admits only a few natural implementations, which leaves open what a match actually shows.

We investigate generated-reference matching using 29{,}970 student submissions from ten Python labs offered in 2021, 2023, and 2025, together with 90{,}000 solution attempts generated retrospectively by three frontier LLMs. We validate the generated solutions using hidden instructor tests, compare code with MOSS after excluding the starter code, and examine the exact abstract syntax tree (AST) forms of selected functions.

The models usually produced correct solutions and, across most assignments, converged on similar implementations. Student submissions matched the generated references more often in later cohorts, including among submissions that passed every hidden test. On tightly specified functions, the models converged on a few exact abstract-syntax-tree forms, and the number of distinct student forms also declined across cohorts, whereas open-ended functions remained diverse in both sources. Most reported overlaps were short, making the minimum match length an important choice when reviewing students' code.

Finally, we discuss how instructors can build a reference bank of generated solutions before releasing an assignment to identify tasks on which generated solutions converge, decide how much review a match warrants, and redesign tasks to elicit tests, reasoning, and intermediate work. These findings support tracking population-level changes in submitted code, while attributing AI use to an individual submission would require additional evidence about how it was produced, such as prompts, revisions, intermediate code, and student disclosures.
\end{abstract}



\maketitle

\input{sections/01-intro}

\input{sections/02-related-work}

\input{sections/03-method}

\input{sections/04-result}

\input{sections/05-discussion}

\input{sections/06-limitation}

\input{sections/07-conclusion}

\input{sections/98-ack}

\appendix
\input{sections/99-appendix}

\balance
\bibliographystyle{ACM-Reference-Format}
\bibliography{references}

\end{document}

%% file: sections/01-intro.tex
\section{Introduction}
\label{sec:intro}

Generative AI has made plausible code cheap to produce. Programming instructors, therefore, need a practical way to decide which submissions warrant closer review. One option is an automated classifier that predicts whether a program is human- or AI-generated. Purpose-built code classifiers can be accurate on code that resembles their training data~\cite{nguyen2024gptsniffer}, yet evaluations show that results degrade across generators, languages, and domains~\cite{orel2025droid,orel2026aicd}, and even under modest code transformations~\cite{pan2024detectors}. Thus, the reliability of this approach depends on how closely the code being checked resembles the code the classifier was trained on.

An alternative approach reuses a familiar workflow from plagiarism detection. First, generate several solutions for an assignment using large language models (LLMs), add them to a reference bank, and compare student code against that bank with a similarity detector such as MOSS~\cite{aiken2026moss}. Prior studies of these generated ``pseudo-submissions'' report promising results in bounded settings~\cite{jegourel2024sieving,bashir2025pseudo,karnalim2026detecting}. Once the reference bank and preprocessing choices are fixed, the comparison returns matching code snippets for inspection. An analyst can then apply a minimum match length to control which reported matches are retained~\cite{schleimer2003winnowing,aiken2026moss}. We call this \emph{generated-reference matching}.

However, a match still does not prove AI use. Generation is stochastic, so a sampled reference bank cannot contain all possible solutions to an assignment. Short CS1 tasks also constrain the available implementations, allowing students and LLMs to converge independently. Finally, the number of matches may change with the minimum length of the code snippet. These uncertainties matter differently at two levels. For an individual submission, a match supports attribution only if it reliably distinguishes AI-produced from independently written code. For a cohort, matching can describe how submitted code changes across tasks and years without attributing a source to any submission. We therefore evaluate generated-reference matching as a population-level measurement.

In this work, we investigate generated-reference matching using three years of CS1 Python lab submissions. We first evaluate whether the LLMs generate correct reference solutions, grading every attempt with the course's hidden instructor tests, and how often the student-visible checker triggers a revision (RQ1). We then compare overlap and exact code forms across LLMs, tasks, and student cohorts (RQ2). Finally, we examine how the minimum match length affects file coverage and which claims retained matches can support (RQ3).

\begin{itemize}
  \item[\textbf{RQ1}:] How functionally correct are frontier-LLM solutions to CS1 assessments, and how often does the student-visible checker trigger a revision?
  \item[\textbf{RQ2}:] How do MOSS overlap and exact code forms vary across LLMs, tasks, and student cohorts?
  \item[\textbf{RQ3}:] Which detection claims can MOSS matching against generated references support on short CS1 tasks, and how does minimum match length affect file coverage and interpretation?
\end{itemize}
\label{sec:rqs}

We study ten Python labs from a CS1 course offered in 2021, 2023, and 2025. The corpus contains 29,970 final student submissions and 90,000 reference attempts generated retrospectively from the corresponding year's materials by three frontier LLMs (GPT-5.5, Gemini~3.1~Pro, and Claude Opus~4.8). Using three providers lets us examine whether convergence extends across models. We apply the same 2026 models and generation protocol to every cohort to measure retrospective resemblance under a consistent protocol (Section~\ref{sec:generation}). We grade every submission file with the hidden instructor tests for its year, compare the code with MOSS after excluding starter code, and count exact abstract syntax tree (AST) forms for five selected functions. 

We use 2021 as a historical reference before the broad adoption of code-generating tools. The offering predates ChatGPT's November 2022 public release and GitHub Copilot's June 2022 general availability. Copilot was available during 2021 only through a limited-seat technical preview~\cite{openai2022chatgpt,github2021copilot,github2022copilotga}.

This study contributes a large-scale empirical characterization of generated-reference matching in CS1. It validates reference correctness and measures checker-guided revisions, compares MOSS overlap across models and cohorts, examines exact forms for five functions with full-pass sensitivity analyses, and quantifies how minimum match length changes coverage. Together, these results show how reference correctness, assignment structure, and matching criteria shape interpretation. They support population-level monitoring and assignment review while identifying the process evidence needed to evaluate individual attribution.

%% file: sections/02-related-work.tex
\section{Related Work}
\label{sec:related-work}

\subsection{Correctness and Revisions in Generated Code}

Evaluations on authentic course material quickly established that code LLMs can solve introductory exercises. Codex produced passing solutions for many CS1 and CS2 problems~\cite{finnieansley2022robots,finnieansley2023cs2}, natural-language revisions repaired many initially unsuccessful Copilot solutions~\cite{denny2023conversing}, and GPT-4 passed varied assessments across multiple programming courses~\cite{savelka2023thrilled}. Consequently, syntactic and functional correctness alone provide little evidence of how a solution was constructed~\cite{becker2023programming,denny2024computing,prather2023robots}.

Students use GenAI for debugging, explanation, generation, and tool learning, with attitudes varying by experience and by whether programming is the learning objective~\cite{keuning2024perceptions}. Immediate performance and learning nevertheless remain distinct. Controlled and in-situ studies report faster or more successful code production~\cite{kazemitabaar2023novice,shihab2025copilot}, but also show that interaction strategy matters: single-prompt solution use can accompany weaker subsequent modification performance~\cite{kazemitabaar2023use}, novices can be led astray by suggestions~\cite{prather2023weird}, and beginning programmers struggle to express intent and verify generated code~\cite{nguyen2024misread}. Better submitted code, therefore, does not directly establish stronger underlying knowledge. Classroom deployments have also explored assistants that provide useful guidance without directly revealing solutions. CodeAid was deployed in a large programming course with constraints intended to balance student help with educator concerns~\cite{kazemitabaar2024codeaid}. Assessment guidance accordingly emphasizes alignment with learning outcomes, comprehension, testing, debugging, explanation, and process evidence~\cite{mahon2024guidelines,denny2024computing,prather2023robots}.

\subsection{Code Similarity Across LLMs, Tasks, and Cohorts}

Short programming tasks can yield structurally similar solutions even when developed independently~\cite{luxtonreilly2013differences}. Required signatures, starter files, construct restrictions, and common idioms can narrow the available forms, while more open tasks may admit many implementations. Prior work has therefore used structural similarity to cluster students' programming solutions~\cite{artser2024clustering}. Other work identifies generated code through one-to-one sample matching~\cite{karnalim2026detecting} or anomaly-feature matching~\cite{karnalim2024detecting}. This motivates measuring both MOSS overlap and exact solution forms across multiple assignments.

Evidence across cohorts remains sparse: an unchanged assignment showed post-ChatGPT changes in submitted code and revision behaviour, but the comparison could not identify individual use~\cite{zhang2026changes}. Cohorts may also differ in instruction, composition, performance, assignment details, shared resources, and available tools.

\subsection{Code Similarity for Retrieval and Authorship Claims}

Source-code similarity systems retrieve overlap, and authorship remains a separate judgement. Winnowing supports scalable fingerprinting~\cite{schleimer2003winnowing}, and MOSS explicitly requires human interpretation because similarity is not itself evidence of plagiarism~\cite{aiken2026moss}. Generated code sharpens this distinction: independent solutions may converge because a task admits few natural implementations, while sampling, editing, or decomposition may suppress overlap. Consistent with both possibilities, generated solutions have evaded suspicious MOSS matches~\cite{biderman2022fooling}, whereas task-specific clustering, pseudo-submission matching, and task-conditioned classifiers have reported strong separation in their evaluated settings~\cite{jegourel2024sieving,bashir2025pseudo,ashkenazi2025task}.

Machine-learning and NLP research similarly show that detection accuracy is conditional. Curvature-based detection and watermarking can perform well when the required model probabilities or generation-time controls are available~\cite{mitchell2023detectgpt,kirchenbauer2023watermark}. Generation-time watermarking has even reached production scale, with SynthID-Text deployed in Gemini and open-sourced for other developers~\cite{dathathri2024synthid,google2026synthid}. The assumptions still bind, however: a watermark presumes a cooperating generator, is less effective on low-entropy output, and loses detector confidence when text is thoroughly rewritten or translated~\cite{dathathri2024synthid,google2026synthid}. Paraphrasing alone likewise substantially degrades trained classifiers, zero-shot methods, and watermark-based approaches~\cite{sadasivan2025reliable,krishna2023paraphrasing}, and broader stress tests that add editing, prompting, and co-generation attacks report similar failures~\cite{wang2024stumbling,shi2024redteaming}. False positives can also be distributed inequitably across human populations~\cite{liang2023bias}. Code-specific classifiers likewise achieve high accuracy when training and evaluation contexts are aligned~\cite{nguyen2024gptsniffer,orel2025codet}. In computing education, a comparison of eight public detectors found lower accuracy for code, non-English text, and paraphrased responses, along with false-positive concerns~\cite{orenstrakh2024detecting}. A later education-focused comparison found poor discrimination across five detectors and several code variants~\cite{pan2024detectors}, while broader evaluations report pronounced failures across unseen languages, domains, LLMs, and human--AI or adversarial code~\cite{orel2025droid,orel2026aicd,orel2026semeval}. Authorship detection can succeed in bounded settings, yet high in-domain accuracy alone does not justify use as an educational instrument.

Taken together, this literature motivates examining generated-reference matching across assignments and student cohorts, where task constraints and independent convergence can shape code similarity. This requires attention to the correctness of generated references, the overlap among generated and student solutions, and the matching criteria that determine which submissions are retrieved. Our study examines these questions to assess what generated-reference matching can reveal about population-level patterns in submitted code.

%% file: sections/03-method.tex
\section{Method}
\label{sec:method}

We retrospectively studied ten CS1 labs across three course offerings, pairing final student submissions with solutions generated by three frontier LLMs for the corresponding year's assignments. RQ1 evaluates correctness and checker-guided revisions. RQ2 compares MOSS overlap and exact AST forms. RQ3 measures coverage at increasing minimum match lengths. Every file is graded with its year-specific instructor tests, and the corresponding starter code is excluded from MOSS.

\subsection{Course and Student Submissions}
\label{sec:corpus}

The study uses ten weekly Python lab assignments (Labs~1--10) from a 12-week introductory programming course (CS1) at a large research-intensive university. Each lab provides a handout, starter code, and a small student-visible checker. Grading uses a larger hidden instructor test suite. The labs contain one to five functions and cover arithmetic, Boolean expressions, strings, lists, loops, functions, data structures, file I/O, and regular expressions. For example, Lab~2 implements constrained logic functions, Lab~5 manipulates strings with nested loops, Lab~8 reads and transforms elevation maps, and Lab~10 finds an email with a regular expression and asks students to write tests. We compare tasks descriptively.

We use the final submitted file per student per lab from the 2021, 2023, and 2025 offerings: 9{,}735, 10{,}987, and 9{,}248 graded submissions, respectively. We use these three offerings because the same lead course instructor designed and ran them with the same material, labs, textbook, and delivery mode. The 2022 and 2024 offerings were taught by different teaching teams and differed too much for comparison. Within the three offerings, the number of sections and the other course instructors varied, about 35\% of teaching assistants continued from one offering to the next, and course tests used different questions of the same types. Labs~2--10 retained the same required function set and core specifications. The 2023 and 2025 student-visible checker assertions were identical, while the 2021-to-later changes were primarily added doctest checks and a revised Lab~8 I/O harness. The hidden suites retained the same core behaviours with a small number of added or revised edge cases. Each file was evaluated against its own year's materials and tests. Student--LLM MOSS analyses exclude four unreadable or non-text files (three from 2023 and one from 2025). The correctness analysis retains all 29{,}970 graded submissions.

Course policy on AI use also differed across cohorts. The 2021 syllabus predated broad access to code-generating LLMs and contained no AI-use policy. The 2023 and 2025 syllabus carried materially identical policies: students could use AI tools for learning course material, practising programming, and coding support, but graded submissions had to be the original work of the individual student, and AI assistance was prohibited on tests and exams. The Lab~4 and Lab~6 handouts in both later offerings also provided an institution-hosted generative-AI tutor QuickTA \cite{kumar2023quickta} for problem-solving guidance while explicitly disallowing direct solutions and requiring students to create and understand submitted work. Thus, access to AI support changed across the observed period, even though the original work expectation for graded labs was stable in 2023 and 2025.

\emph{Ethics.} The study was approved under the university's Research Ethics Board (\#38966). Submissions were de-identified before analysis. We report only aggregate statistics, and no similarity result is used to make or imply a misconduct determination about any student.

\subsection{Generating LLM Solutions}
\label{sec:generation}

\subsubsection{LLMs and samples}
We study three frontier proprietary LLMs: GPT-5.5, Gemini~3.1~Pro, and Claude Opus~4.8, accessed through their providers' batch APIs. Within each of the three major provider families, we selected the most capable coding model accessible at the time of the study. The cross-LLM comparison examines whether convergence extends across models. Providers routinely deprecate and sunset earlier models, preventing us from reproducing the model set available to students in 2023 and 2025~\cite{openai2026deprecations, anthropic2026deprecations, google2026modellifecycle, google2026geminimodels}. We therefore use the same frontier model set for all three cohorts to provide a consistent retrospective reference. All models used an explicit 24{,}000-token output limit and provider-default temperature/top-$p$ (1.0/0.98 for GPT-5.5, 1.0/0.95 for Gemini, and 1.0/0.99 for Claude). We initiated 1{,}000 independent attempts per LLM, lab, and cohort year (30{,}000 per LLM, 90{,}000 total). Generation settings are held constant across cohorts while assignment inputs reflect each year's materials.

The generation run consumed 491.8 million tokens and cost approximately US\$4{,}284 at provider batch rates (4.8 cents per attempt).

\subsubsection{Prompts and revisions}
Each generation attempt receives the complete year-specific lab handout (converted from the original PDF to Markdown), the unmodified starter file, and a request for a completed solution file. We run the result against only the student-visible checker. If that checker fails, the LLM receives its output along with the original handout and starter code and may revise the file. It can receive at most two such revision prompts (three prompts total). Generation stops when the student-visible checker passes or the second revision is exhausted. Hidden instructor tests are never included in a generation prompt and are run only during retrospective grading. See Appendix \ref{app:prompt-initial} for the initial generation prompt.

For RQ1, we record first-prompt success against the visible checker, the number of revision prompts, and whether the attempt remains unresolved after the final revision.

\subsection{Generated-Solution Correctness and Revisions}
\label{sec:grading}

Each student submission and LLM attempt is placed in a fresh working directory and run against the hidden instructor pytest suite for its cohort year and lab, with a 60-second timeout to catch non-terminating code. For each file, we report the mean test score (passed instructor tests divided by total tests) and whether it passes the full suite. We call a file that passes every hidden instructor test a \emph{full-pass} file. The MOSS sensitivity analysis restricts student submissions to full-pass files while retaining all generated attempts in the reference bank. The exact-form sensitivity analysis likewise uses full-pass student files. A file that crashes before pytest reports test counts receives a score of zero.

\subsection{Measuring Code Similarity with MOSS}
\label{sec:moss}

We run MOSS locally so that student code remains on institutional infrastructure~\cite{schleimer2003winnowing,aiken2026moss}. In both designs below, we use Python mode (\texttt{-l python}) and register the corresponding starter file as a base file (\texttt{-b}) so distributed code is excluded from matching. Controlled runs of two unchanged starter files produced no reported match.

\subsubsection{Similarity among LLM solutions}
For each cohort and lab, we compare all 1{,}000 attempts from each of the three LLMs (3{,}000 files). We label a returned pair as within-LLM when both files came from the same model, or cross-LLM when they came from different models, and divide the number of matches by the number of pairs that could have been formed. Each comparison contains 3{,}000{,}000 possible cross-LLM pairs and 1{,}498{,}500 possible within-LLM pairs. The within-LLM rate provides a baseline for interpreting the cross-LLM rate. We set the maximum-occurrence parameter above the 3{,}000-file pool so that MOSS does not discard code simply because many generated answers share it.

The primary LLM--LLM analysis removes module and function docstrings from every attempt and from the starter file before MOSS. Inspection showed that LLMs often added similar conventional doctests that starter-file exclusion could not remove. Removing them, therefore, gives a more conservative comparison of implementation code. The student--LLM analysis retains docstrings because it compares files as submitted or generated. Rates from the two MOSS designs are therefore not directly comparable.

\subsubsection{Student--LLM matching across cohorts}
For every LLM, cohort, and lab, we submit all available student files and all 1{,}000 LLM attempts to one MOSS run. We set MOSS's maximum-occurrence parameter (\texttt{-m}) to $250$, suppressing code that appears in more than roughly one quarter of the average lab--cohort student pool\footnote{This occurrence filter is distinct from the minimum match length varied in RQ3.}. A \emph{cross-match} is a returned pair containing one student file and one LLM attempt. We report four measures. Pair-level incidence is the number of cross-matches per 100{,}000 possible student--LLM pairs. Student-file coverage is the number of distinct student files with a match per 1{,}000 student files. LLM-attempt coverage is the percentage of attempts with a student match. The fourth measure is the MOSS-reported matched-line count for each pair.

This analysis retains docstrings in student and LLM files, matching the code as submitted or generated. Starter-file exclusion removes distributed text from initiating a match, but a reported MOSS code snippet can span nearby retained documentation once a genuine code match exists. We therefore interpret the matched-line count as the size of the MOSS-reported code snippet, which may include retained documentation and code that students write in similar ways for many legitimate reasons~\cite{simon2020choosing}.

For the cohort comparison in RQ2, we compare the 2023 and 2025 measures with the 2021 historical reference and retain the lab-specific results. We repeat incidence and student-file coverage using only full-pass submissions to examine whether the longitudinal pattern persists among files that pass every instructor test.

\subsubsection{Varying the minimum match length}
Starting from the same student--LLM match lists, we vary the minimum match length in MOSS-reported lines. We apply this post-processing filter to returned pairs while keeping the MOSS matching algorithm unchanged. At each threshold, a student file or LLM attempt is counted as matched if it has at least one cross-match meeting that threshold. We report the share of student files and LLM attempts retained by LLM, cohort, and threshold. This analysis quantifies how strongly candidate coverage depends on the chosen minimum match length.

\subsection{Counting Exact Code Forms}
\label{sec:diversity}

As a complementary analysis of exact code forms, we extract the five function bodies listed in Table~\ref{tab:selected-functions} from Labs~2 and~5 and serialize their Python ASTs after removing a leading docstring.

We selected these five functions to test whether longitudinal convergence appears consistently across different degrees of implementation freedom or whether exceptions emerge. Four tasks channel correct solutions toward a small set of intuitive implementations, whereas \texttt{loopy\_madness} leaves more choices about traversal, state, and string construction~\cite{simon2020choosing}. Because all five are named standalone functions from Labs~2 and~5, whose required functions and specifications were the same in all three offerings, they permit direct comparison of the same tasks across cohorts. Eligible files must parse and contain the named function. Identifier names and all other body syntax remain part of the representation. Two bodies count as the same form only when these serialized ASTs are equal.

We use the five functions for two RQ2 comparisons. For the student--LLM comparison, we compare 2021 students with the pooled LLM attempts generated for the corresponding 2021 task. Because the number of observed forms increases with sample size, we use analytical rarefaction to reduce the larger LLM pool to the corresponding student-pool size. Each reported LLM value is the exact expected number of forms in a sample of that size, computed from the full frequency distribution without random draws. The student value uses the full eligible pool. We express both counts per 1{,}000 files and do not attach Poisson intervals to this equal-size comparison because distinct forms are dependent, saturating counts.

For the cohort comparison, we trace the number of distinct student forms across 2021, 2023, and 2025 using each cohort's full eligible pool. We report the distinct-form rate with a Poisson 95\% confidence interval and an exact 95\% interval on the 2021-to-2025 rate ratio. These intervals are descriptive approximations because distinct forms are dependent, saturating counts. As a correctness sensitivity analysis, we then restrict to full-pass files and analytically rarefy all three cohorts to the smallest eligible full-pass pool for that function. This separates the number of distinct forms from the unequal number of correct submissions across years.

\subsection{Statistical Analysis and Scope of Claims}
\label{sec:stats}

For correctness and revisions (RQ1), we report observed means, full-pass proportions, checker outcomes, and their denominators. Mean instructor-test scores have two-sided 95\% $t$ intervals computed over submission-level student scores or attempt-level LLM scores.

For MOSS similarity, cohort change, and minimum match length (RQ2--RQ3), we report counts, normalized rates, coverage, rate or coverage ratios, and matched-line distributions. Figure~\ref{fig:moss-fig01-incidence-coverage} uses exact Poisson intervals for cross-match incidence and Wilson intervals for student-file coverage. Cohort annotations compare 2023 with 2021, 2025 with 2021, and 2025 with 2023 using Poisson log rate-ratio Wald tests for incidence and two-proportion $z$ tests for coverage. Benjamini--Hochberg correction is applied within each measure family. A file or LLM attempt can appear in many MOSS pairs, and students contribute files to multiple labs. These dependencies can make the intervals and tests overstate precision. We therefore treat them as model-based descriptive annotations and do not attach significance tests to the matched-line distributions or threshold curves.

For the exact-form distribution comparisons, we use full eligible-file counts to construct source-by-form contingency tables, pool forms with fewer than five observations across the compared groups, and apply Pearson's $\chi^2$ test of independence. Within each five-function comparison family, we control the false discovery rate using the Benjamini--Hochberg procedure and report adjusted $q$-values along with Cram\'er's~$V$. These tests assess the association between source or cohort and the exact form within the five selected functions.

Across all analyses, the unit and denominator are stated with each result. The analyses measure generated-solution correctness, code similarity, changes across cohorts, and match coverage. Verified student source labels are unavailable, so detector accuracy and individual AI-use attribution remain outside the study's scope.

%% file: sections/04-result.tex
\section{Results}
\label{sec:results}

\subsection{RQ1: Generated-Solution Correctness and Revisions}
\label{sec:results-rq1}

\input{tables/koli2026_mean_scores}

Across LLM--year combinations, with labs pooled within each combination, mean instructor-test scores ranged from 98.93\% to 99.99\%, compared with 76.37\%, 80.76\%, and 88.03\% for student submissions in 2021, 2023, and 2025, respectively (Table~\ref{tab:mean-scores}). Pooled the same way, LLM full-pass rates ranged from 86.97\% to 99.95\%. Corresponding student rates were 39.96\%, 44.90\%, and 61.71\%. Most generated solutions passed the instructor tests, although performance varied by lab.

Most generated solutions did not need a checker-guided revision: first-prompt success against the visible checker ranged from 96.61\% to 99.98\% across LLM--year combinations. Only Gemini left attempts unresolved after two permitted revisions, at 1.71\% in 2023 and 1.98\% in 2025. Passing the student-visible checker did not always mean passing every instructor test. The largest full-pass gap occurred on Lab~4 in 2021: Gemini and Claude passed the student-visible checker on the first prompt in 100.0\% and 99.9\% of attempts, respectively, but passed the full hidden suite in only 0.4\% and 0.0\% (4 and 0 of 1{,}000). Their mean hidden-test scores were nevertheless 92.34\% and 92.31\%.

The instructor tests characterize reference correctness. We next examine whether matches are distinctive by measuring convergence across models and tasks.

\subsection{RQ2: Similarity Across LLMs, Tasks, and Student Cohorts}
\label{sec:results-rq2}

\subsubsection{Cross-LLM and Task-Level Similarity}
\label{sec:results-rq2-convergence}

Solutions from different LLMs matched broadly and consistently across most lab exercises. Using all 1{,}000 attempts from each LLM and pooling the ten labs within each cohort, 87.79--88.14\% of cross-LLM pairs matched after docstring removal. The corresponding within-LLM rates were 91.35--91.72\%, only 3.40--3.61 percentage points higher (Table~\ref{tab:model-convergence}). Across Labs~1--9, cross-LLM rates were 95.48--96.08\%. On Lab~10 they were 13.18--22.11\%. The Lab~10 task helps explain this difference: students implement one short regular-expression function and submit their tests in a separate file, while only the implementation file entered the MOSS corpus. Correct answers can compress into a single dense regex, and equivalent regexes can use different syntax, leaving little shared code for MOSS to match.

\input{tables/koli2026_functions_description}
\input{tables/koli2026_model_convergence}

The exact-form analysis made the task contrast concrete. At the constrained endpoint, \texttt{my\_and} asks students to express a Boolean conjunction using only \texttt{not} and \texttt{or}, directing solutions toward De Morgan's law. All three LLMs produced the same exact AST form. At the open-ended endpoint, \texttt{loopy\_madness} asks students to interweave two strings while traversing the shorter string backward and forward, leaving choices about indices, direction state, and string construction. The 3{,}000 LLM attempts produced 2{,}288 forms, with none shared by all three models.

\begin{figure}[t]
  \centering
  \includegraphics[width=\linewidth]{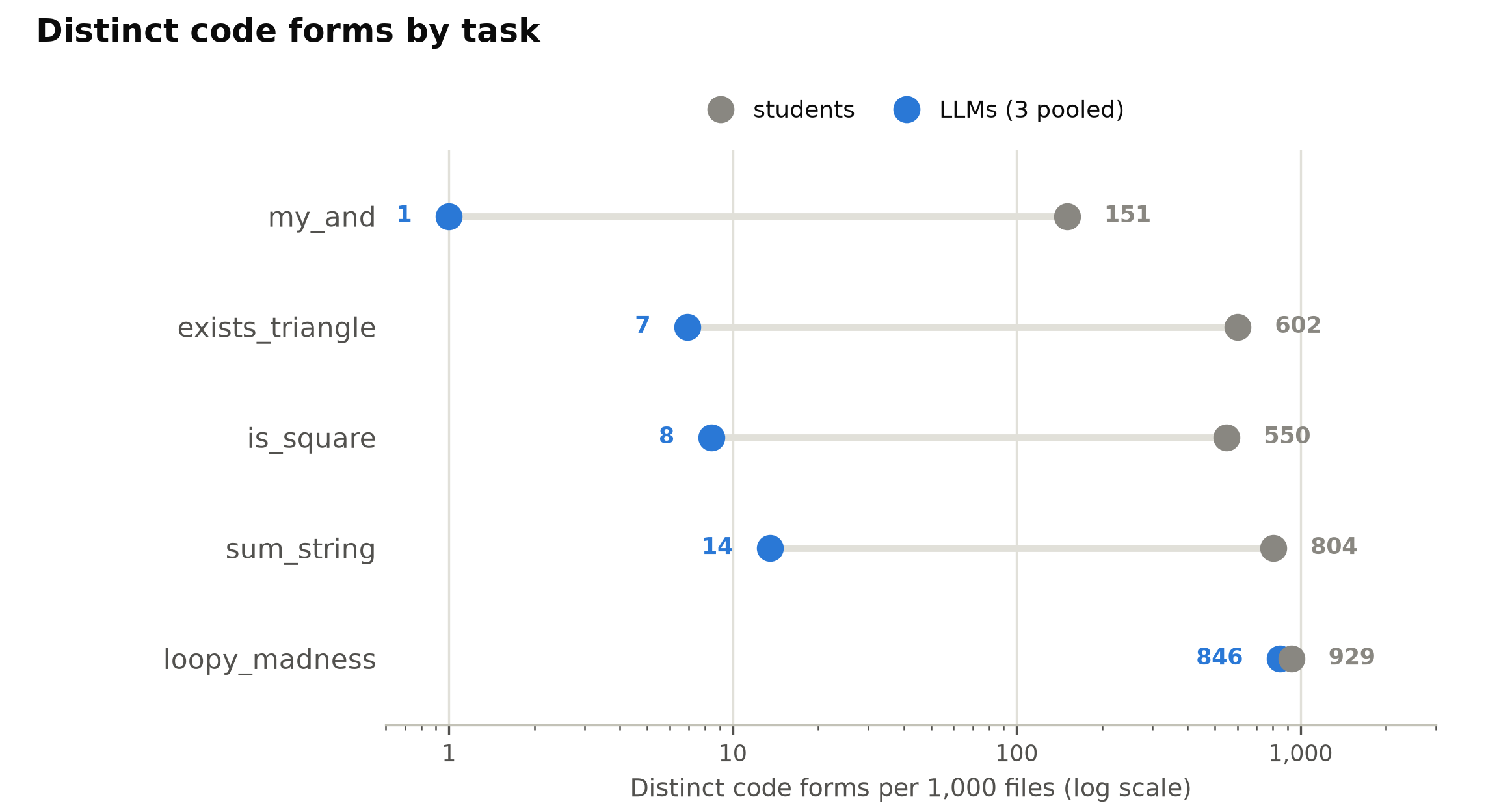}
  \Description{A logarithmic dumbbell plot compares exact AST-form counts for 2021 students and pooled frontier-LLM attempts across five selected functions. Frontier-LLM counts are much smaller for four functions and similar to the student count for the fifth.}
  \caption{Distinct exact AST forms per 1{,}000 eligible files. Student values use all eligible files. LLM values are exact expected counts at the student-pool size, computed from all LLM files without random sampling.}
  \label{fig:moss-fig08-implementation-diversity}
\end{figure}

This contrast extended across the five selected functions. On the first four, tightly specified functions in Table~\ref{tab:selected-functions}, the pooled LLM attempts yielded 1--14 expected exact AST forms per 1{,}000 files after docstring removal and rarefaction to the student-pool size, compared with 151--804 among 2021 student submissions. On \texttt{loopy\_madness}, the corresponding rates were 846 and 929 per 1{,}000 (Figure~\ref{fig:moss-fig08-implementation-diversity}). Source and exact form were associated for all five functions (Pearson's $\chi^2$, BH-adjusted $q<.001$), with Cram\'er's~$V$ ranging from 0.30 for \texttt{loopy\_madness} to 0.96 for \texttt{is\_square} and \texttt{sum\_string}.

\begin{figure*}[t]
  \centering
  \includegraphics[width=\textwidth]{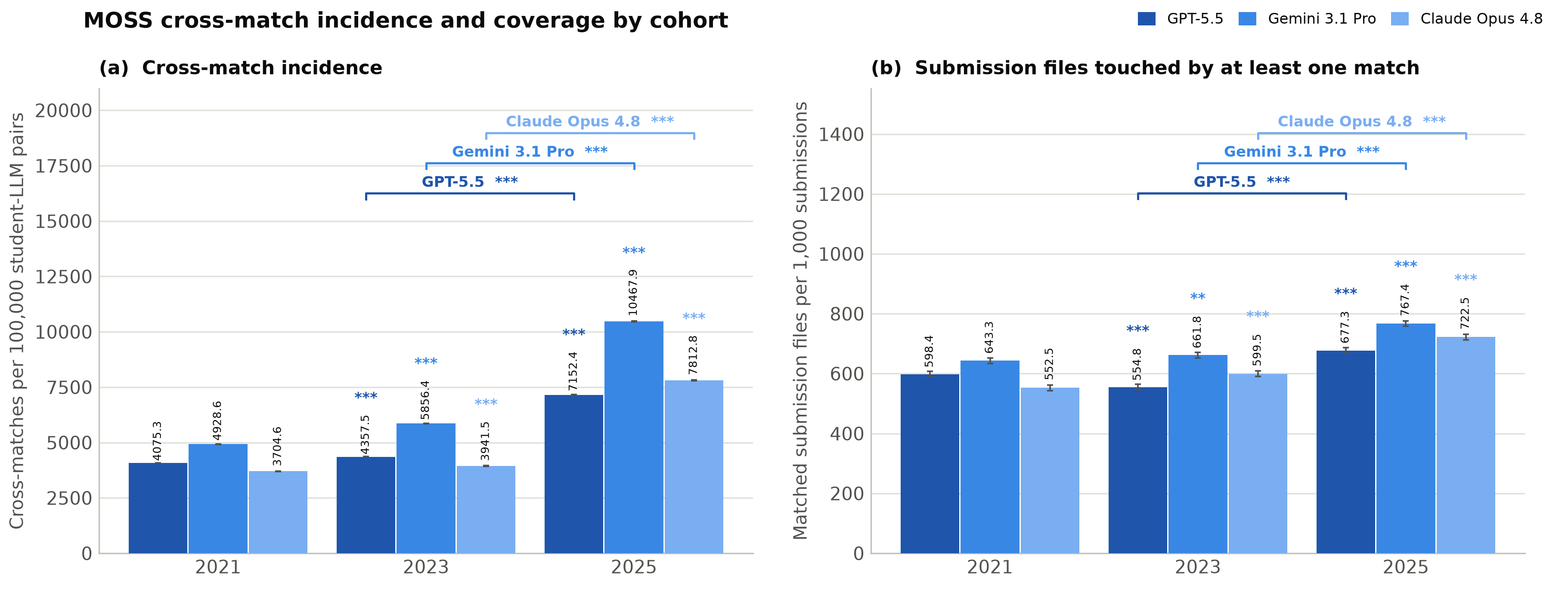}
  \Description{Two grouped bar charts show student--LLM cross-match incidence and student-file match coverage for GPT-5.5, Gemini 3.1 Pro, and Claude Opus 4.8 in 2021, 2023, and 2025. Both measures are higher in 2025 than in the earlier cohorts for all three models.}
  \caption{Student--LLM MOSS matching by cohort: (a)~cross-matches per 100{,}000 possible pairs and (b)~matched submissions per 1{,}000. Error bars show 95\% intervals. Stars above each 2023 and 2025 bar compare that cohort with 2021, and brackets compare 2025 with 2023, using Benjamini--Hochberg adjusted $q$-values (* $q<.05$, ** $q<.01$, *** $q<.001$).}
  \label{fig:moss-fig01-incidence-coverage}
\end{figure*}

\begin{figure*}[t]
  \centering
  \includegraphics[width=\textwidth]{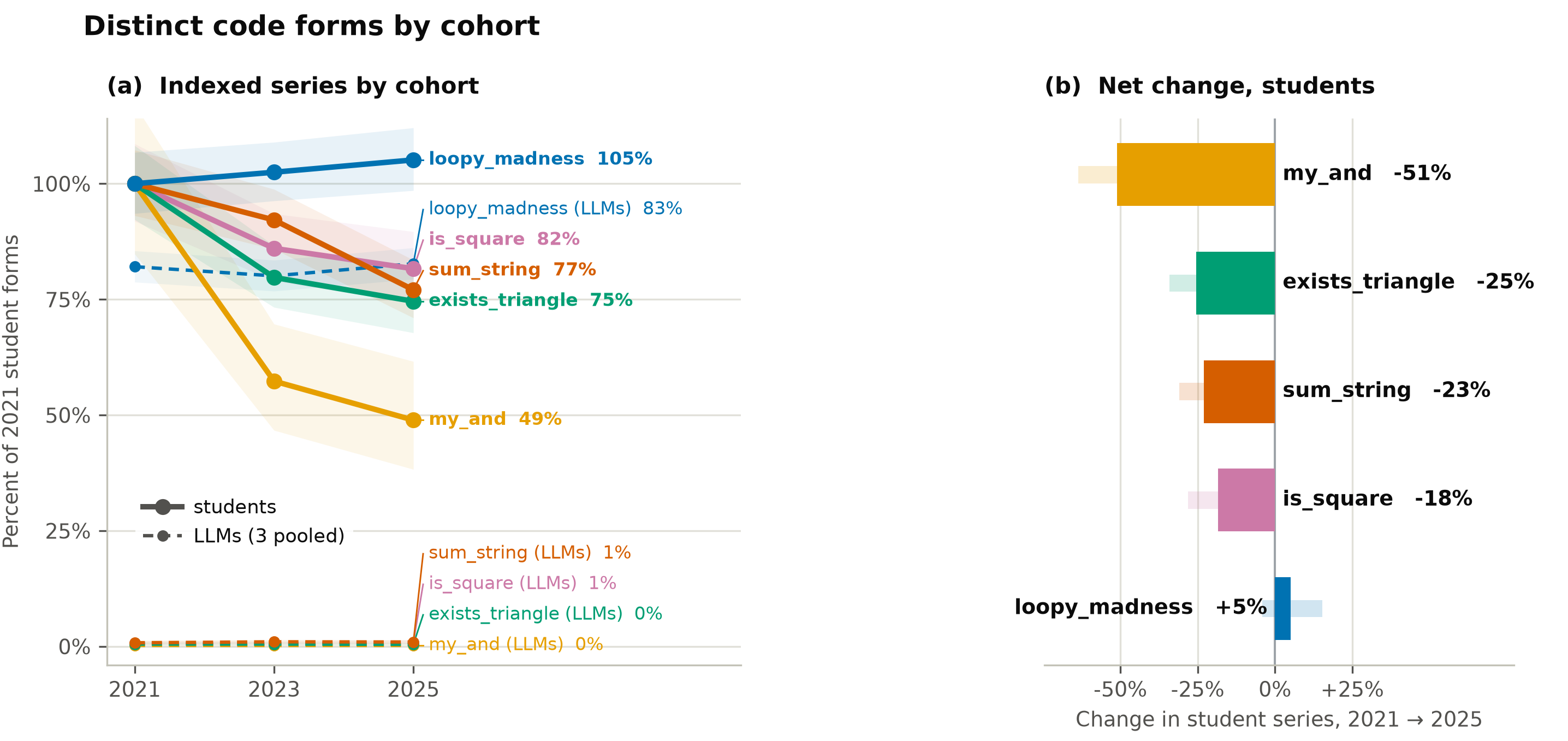}
  \Description{Indexed line plots show the number of distinct exact student AST forms from 2021 through 2025 for five selected functions, with pooled frontier-LLM attempts as a comparison. A companion bar chart shows the net change in students from 2021 to 2025.}
  \caption{Distinct exact AST forms per 1{,}000 files by cohort for five selected functions. Panel~(a) indexes student and pooled-LLM series to the 2021 student value. Panel~(b) shows the 2021--2025 change in student forms. Shading shows Poisson 95\% intervals for form rates. Bars show 95\% intervals derived from the 2025/2021 rate ratio, accounting for both cohorts under that model. Full cohorts include functionally incorrect submissions.}
  \label{fig:moss-fig12-diversity-trend}
\end{figure*}

LLMs therefore arrived at similar code across most assignments. The Lab~10 exception shows why per-task baselines remain necessary.

Having established this broadly convergent but task-dependent baseline, we next examine whether student code moved toward the generated reference bank across cohorts.

\subsubsection{Changes Across Student Cohorts}
\label{sec:results-rq2-cohort}
Student--LLM match incidence was 1.06--1.19 times its 2021 rate in 2023 and 1.76--2.12 times that rate in 2025 (Figure~\ref{fig:moss-fig01-incidence-coverage}). Every annotated incidence comparison had BH-adjusted $q<.001$, including the small 2023 shifts. The change also varied by assignment (Table~\ref{tab:lab-year-rates}). When matches shorter than 30 lines were excluded, the 2025/2021 incidence ratios were 1.59 for GPT-5.5, 2.30 for Gemini, and 5.67 for Claude, while student-file match coverage rose from 29.2--35.4 per 1{,}000 in 2021 to 80.2--136.6 per 1{,}000 in 2025.

The increase persisted when only full-pass student submissions were included. Among full-pass submissions, incidence per 100{,}000 pairs rose from 4{,}774.8 to 7{,}579.6 for GPT-5.5, from 5{,}059.8 to 11{,}019.7 for Gemini, and from 3{,}498.1 to 8{,}173.2 for Claude. The corresponding 2025/2021 ratios were 1.59, 2.18, and 2.34. Rising student correctness alone, therefore, does not explain the direction of the observed student--LLM similarity shift.

\input{tables/koli2026_lab_year_rates}

Student solutions also became less varied in four of the five selected functions. Relative to 2021, the number of exact forms per 1{,}000 files in 2025 was lower by 51.1\% for \texttt{my\_and} (95\% CI 35--64\%), 25.4\% for \texttt{exists\_triangle} (16--34\%), 22.9\% for \texttt{sum\_string} (14--31\%), and 18.4\% for \texttt{is\_square} (7--28\%). It was 5.1\% higher for \texttt{loopy\_madness} (95\% CI $-4\%$ to $+15\%$), as Figure~\ref{fig:moss-fig12-diversity-trend} shows. The four declines exclude zero, whereas the \texttt{loopy\_madness} interval does not. These comparisons use each function's full cohort and include submissions that are functionally incorrect. In the full-pass comparison, rarefied to a common sample size within each function, the four constrained functions still declined by 14.4--32.1\% from 2021 to 2025, whereas \texttt{loopy\_madness} changed by $+2.3\%$.

For additional context on student performance, final-exam mean (median) scores were 55.3 (56.7) in 2021 ($N=990$), 54.6 (55.7) in 2023 ($N={}$1{,}002), and 49.8 (48.8) in 2025 ($N=998$). The exams assessed the same course topics using different questions.

Together, the two analyses show population-level change: student code matched the generated reference bank more often, and four functions appeared in fewer exact forms by 2025. Neither analysis explains why the change occurred nor identifies any student's use of AI.

These population-level changes do not determine how many individual files a detection workflow would retrieve. We therefore next vary the minimum match length and measure the resulting candidate volume.

\subsection{RQ3: Detection Claims at Different Minimum Match Lengths}
\label{sec:results-rq3}

RQ2 established that similarity can be routine on some tasks and unusual on others. RQ3, therefore, asks how student--LLM MOSS matches can function in a detection workflow. We quantify how the minimum match length changes the candidate set and then distinguish what a retained match does and does not establish.

\begin{figure*}[t]
  \centering
  \includegraphics[width=0.9\textwidth]{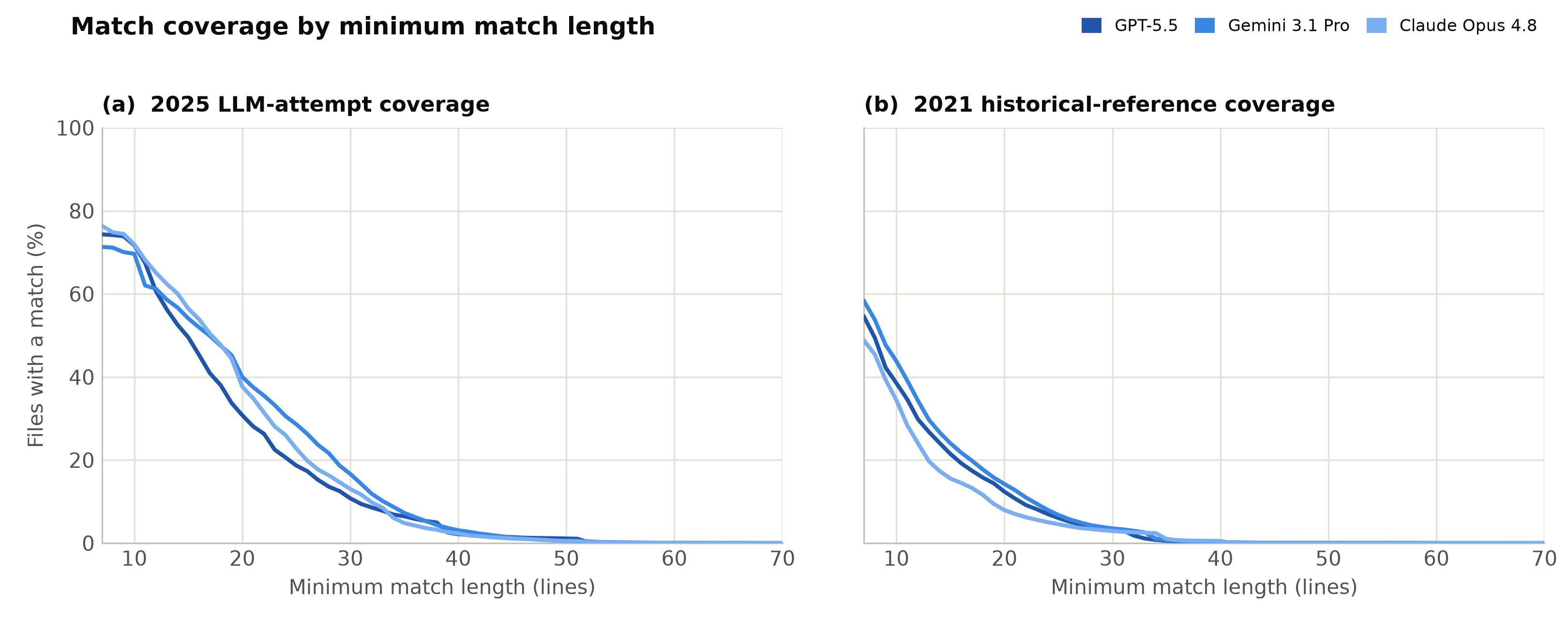}
  \Description{Two line charts show the percentage of files with at least one student--LLM MOSS match as the minimum match length rises. Match coverage falls steeply for 2025 LLM attempts and for 2021 historical-reference student submissions.}
  \caption{Percentage of 2025 LLM attempts (a) and 2021 historical-reference student submissions (b) with at least one student--LLM match at or above each minimum match length. Starter code was excluded through MOSS base files. A match marks code shared between a student submission and an LLM attempt.}
  \label{fig:moss-fig06-threshold-sensitivity}
\end{figure*}

Most reported matches involved short snippets. Across cohorts, the median reported overlap spanned 8--9 lines, and 97.0--99.4\% of cross-matches contained fewer than 30 matched lines.

Raising the minimum match length rapidly reduced coverage for both student files and LLM attempts (Figure~\ref{fig:moss-fig06-threshold-sensitivity}). At 30 lines, 2.9--3.5\% of 2021 historical-reference student files and 10.8--16.7\% of 2025 LLM attempts still had a match. At 50 lines, student-file coverage was near zero, and LLM-attempt coverage fell to 0.5--1.1\%. The minimum match length, therefore, controls how many files remain for review after filtering the MOSS output. These coverage patterns do not reveal which files, if any, were written using AI. That conclusion requires verified source labels or other evidence beyond the match.

%% file: tables/koli2026_mean_scores.tex
\begin{table}[H]
  \caption{Mean instructor-test scores (\%, mean $\pm$ 95\% CI half-width).}
  \label{tab:mean-scores}
  \setlength{\tabcolsep}{3pt}
  \begin{tabular}{@{}lccc@{}}
    \toprule
    \textbf{Source} & \textbf{2021} & \textbf{2023} & \textbf{2025} \\
    \midrule
    Students & $76.37\mathbin{\pm}0.63$ & $80.76\mathbin{\pm}0.53$ & $88.03\mathbin{\pm}0.47$ \\
    GPT-5.5 & $99.29\mathbin{\pm}0.06$ & $99.43\mathbin{\pm}0.05$ & $99.97\mathbin{\pm}0.02$ \\
    Gemini~3.1~Pro & $98.93\mathbin{\pm}0.07$ & $99.82\mathbin{\pm}0.04$ & $99.85\mathbin{\pm}0.03$ \\
    Claude Opus~4.8 & $99.20\mathbin{\pm}0.05$ & $99.99\mathbin{\pm}0.01$ & $99.98\mathbin{\pm}0.02$ \\
    \bottomrule
  \end{tabular}
\end{table}

%% file: tables/koli2026_functions_description.tex
\begin{table}[t]
\caption{The five selected functions used in the exact-form analysis, the
labs they come from, and their specification constraints.}
\label{tab:selected-functions}
\begin{tabular}{@{}cl>{\raggedright\arraybackslash}p{4.8cm}@{}}
\toprule
\textbf{Lab} & \textbf{Function} & \textbf{Task and constraints} \\
\midrule
2 & \texttt{my\_and} & Logical AND of two booleans using only
\texttt{not} and \texttt{or}, with no \texttt{==}, \texttt{bool()}, or
arithmetic operators. \\
2 & \texttt{exists\_triangle} & Return whether three side lengths form a
triangle with positive area, without \texttt{if} statements. \\
2 & \texttt{is\_square} & Test whether an integer is a perfect square
using only arithmetic and comparisons, with no \texttt{if}, \texttt{and},
or \texttt{or}. \\
5 & \texttt{sum\_string} & Alternating digit sum: add odd-positioned
digits, subtract even-positioned digits. \\
5 & \texttt{loopy\_madness} & Interweave two strings, looping
backwards-and-forwards through the shorter string until the longer one
is exhausted. \\
\bottomrule
\end{tabular}
\end{table}

%% file: tables/koli2026_model_convergence.tex
\begin{table}[t]
\caption{Percentage of all possible solution pairs matched by MOSS after docstring removal, using all 1{,}000 attempts per LLM for every cohort and lab. ``Cross'' pairs use different LLMs. ``Within'' pairs use the same LLM.}
  \label{tab:model-convergence}
  \setlength{\tabcolsep}{4pt}
  \begin{tabular}{ccccc}
    \toprule
    & \multicolumn{3}{c}{\textbf{Cross}} & \textbf{Within} \\
    \cmidrule(lr){2-4}\cmidrule(lr){5-5}
    \textbf{Cohort} & \textbf{All labs} & \textbf{Labs 1--9} & \textbf{Lab 10} & \textbf{All labs} \\
    \midrule
    2021 & 87.79 & 96.08 & 13.18 & 91.39 \\
    2023 & 87.94 & 95.96 & 15.81 & 91.35 \\
    2025 & 88.14 & 95.48 & 22.11 & 91.72 \\
    \bottomrule
  \end{tabular}
\end{table}

%% file: tables/koli2026_lab_year_rates.tex
\begin{table}[t]
  \caption{Student--LLM MOSS cross-match incidence per 100{,}000 possible pairs by LLM, lab, and cohort, computed from complete match lists. Cell shading is on a logarithmic scale because rates span nearly three orders of magnitude. The printed values carry the same information.}
  \label{tab:lab-year-rates}
  \footnotesize
  \setlength{\tabcolsep}{2pt}
  \begin{tabular}{rrrrrrrrrr}
    \toprule
    & \multicolumn{3}{c}{2021} & \multicolumn{3}{c}{2023} & \multicolumn{3}{c}{2025} \\
    \cmidrule(lr){2-4}\cmidrule(lr){5-7}\cmidrule(lr){8-10}
    Lab & GPT & Gemini & Claude & GPT & Gemini & Claude & GPT & Gemini & Claude \\
    \midrule
    1 & \cellcolor{blue!31}1,101 & \cellcolor{blue!20}294 & \cellcolor{blue!15}154 & \cellcolor{blue!30}1,060 & \cellcolor{blue!25}572 & \cellcolor{blue!14}133 & \cellcolor{blue!34}1,625 & \cellcolor{blue!21}342 & \cellcolor{blue!27}703 \\
    2 & \cellcolor{blue!9}72 & \cellcolor{blue!20}294 & \cellcolor{blue!35}1,791 & \cellcolor{blue!9}74 & \cellcolor{blue!20}308 & \cellcolor{blue!35}1,943 & \cellcolor{blue!17}204 & \cellcolor{blue!21}350 & \cellcolor{blue!36}2,254 \\
    3 & \cellcolor{blue!48}9,961 & \cellcolor{blue!45}6,814 & \cellcolor{blue!48}9,215 & \cellcolor{blue!46}7,920 & \cellcolor{blue!44}5,750 & \cellcolor{blue!46}7,340 & \cellcolor{blue!48}9,757 & \cellcolor{blue!49}10,826 & \cellcolor{blue!48}9,488 \\
    4 & \cellcolor{blue!30}973 & \cellcolor{blue!36}2,149 & \cellcolor{blue!38}2,878 & \cellcolor{blue!36}2,148 & \cellcolor{blue!38}2,673 & \cellcolor{blue!30}980 & \cellcolor{blue!39}2,938 & \cellcolor{blue!42}4,364 & \cellcolor{blue!40}3,506 \\
    5 & \cellcolor{blue!46}7,329 & \cellcolor{blue!50}11,705 & \cellcolor{blue!41}3,914 & \cellcolor{blue!49}10,334 & \cellcolor{blue!50}11,867 & \cellcolor{blue!45}6,753 & \cellcolor{blue!51}13,338 & \cellcolor{blue!52}16,693 & \cellcolor{blue!48}9,840 \\
    6 & \cellcolor{blue!41}4,160 & \cellcolor{blue!49}11,396 & \cellcolor{blue!47}8,801 & \cellcolor{blue!42}4,625 & \cellcolor{blue!51}13,547 & \cellcolor{blue!45}7,023 & \cellcolor{blue!49}10,230 & \cellcolor{blue!55}23,006 & \cellcolor{blue!50}12,440 \\
    7 & \cellcolor{blue!40}3,487 & \cellcolor{blue!38}2,664 & \cellcolor{blue!25}528 & \cellcolor{blue!35}1,791 & \cellcolor{blue!41}4,262 & \cellcolor{blue!32}1,327 & \cellcolor{blue!41}3,865 & \cellcolor{blue!48}9,410 & \cellcolor{blue!41}4,107 \\
    8 & \cellcolor{blue!47}8,376 & \cellcolor{blue!47}8,146 & \cellcolor{blue!44}5,586 & \cellcolor{blue!48}9,511 & \cellcolor{blue!49}10,332 & \cellcolor{blue!44}5,957 & \cellcolor{blue!52}16,118 & \cellcolor{blue!54}20,572 & \cellcolor{blue!55}21,991 \\
    9 & \cellcolor{blue!43}5,077 & \cellcolor{blue!45}6,505 & \cellcolor{blue!36}2,023 & \cellcolor{blue!44}5,549 & \cellcolor{blue!48}9,763 & \cellcolor{blue!38}2,743 & \cellcolor{blue!50}12,853 & \cellcolor{blue!54}19,696 & \cellcolor{blue!48}9,898 \\
    10 & \cellcolor{blue!28}761 & \cellcolor{blue!4}40 & \cellcolor{blue!37}2,308 & \cellcolor{blue!30}1,008 & \cellcolor{blue!19}273 & \cellcolor{blue!44}5,670 & \cellcolor{blue!35}1,787 & \cellcolor{blue!33}1,451 & \cellcolor{blue!44}5,871 \\
    \bottomrule
  \end{tabular}
\end{table}

%% file: sections/05-discussion.tex
\section{Discussion}
\label{sec-discussion}

Models from different providers repeatedly converged on similar implementations. Later student cohorts matched the reference bank more often and used fewer exact forms on constrained functions. We discuss how task structure and reference correctness shape match interpretation, what the cohort changes support, and how instructors can use these findings.

\subsection{Task Structure and Cross-Model Convergence}
\label{sec-discussion-review}

The first pattern is frequent matching between solutions from different LLMs across most labs. Pooled across all ten labs, solutions from different LLMs matched at rates of about 88\%, only 3.40--3.61 percentage points below the corresponding within-LLM rates. Pooled across Labs~1--9, the cross-LLM rates exceeded 95\%. The exact-form analysis showed the same concentration. For the four constrained functions, the rarefied LLM pool yielded only 1--14 expected forms per 1{,}000 files, compared with 151--804 among 2021 student submissions. This pattern spans providers and is consistent with common influences such as assignment structure and programming conventions.

The assignments help explain the concentration. CS1 tasks fix the function signature, provide starter code, and sometimes prohibit an obvious construct. For example, \texttt{my\_and} has a small set of natural solutions once \texttt{and} is prohibited~\cite{simon2020choosing,luxtonreilly2013differences}. On a task like this, similarity to generated code can simply be similarity to a canonical solution that an independent student could reasonably produce. The reference bank shows how strongly the task channels the sampled models toward a canonical form.

The exceptions highlight the number and location of implementation choices. \texttt{loopy\_madness} permits choices about traversal, state, and string construction throughout the function. Its exact-form rates were high for both sources: 846 expected forms per 1{,}000 files in the rarefied LLM pool and 929 among student submissions. Lab~10 concentrates much of the implementation variation in one compact regular expression. Equivalent regular expressions can encode the same behaviour using different syntax, and within such a short function, those differences leave little shared code for MOSS to match. GPT-5.5 and Claude frequently produced the same expression, whereas Gemini produced a wider range, making the Lab~10 match rate more dependent on the generator's preferred regular-expression syntax. The match rate must therefore be read alongside the form of the expected solution when comparing tasks.

These findings support interpreting matches on a per-task basis. On a tightly constrained task, matching is common because the task channels solutions toward a small set of natural forms. On a more open task, convergence is less routine because the solution space is broader. On a short artifact, the absence of a match may reflect surface-form differences rather than a wider solution space~\cite{aiken2026moss, biderman2022fooling}. Comparisons over time, therefore, require a consistent protocol and attention to the nature of the task.

\subsection{Student Code Moved Toward the Models, and Toward Each Other}
\label{sec-discussion-levels}

By 2025, student--LLM match incidence was 1.76--2.12 times its 2021 level. The increase persisted among full-pass submissions. Independent of MOSS and the reference bank, four constrained functions appeared in fewer exact forms, including after restricting to full-pass files and equalizing sample size. The relatively open-ended \texttt{loopy\_madness} function remained diverse. Together, these analyses document changes in resemblance and diversity that persist among correct submissions.

The timing does not identify a cause. Match incidence in 2023 sat only slightly above 2021, even though general-purpose chat assistants were available throughout that offering, and the course provided an AI tutor in both later years. The much larger increase appeared in 2025. Increased LLM use is consistent with this pattern, but teaching-team composition, permitted support, cohort composition, shared resources, and smaller delivery or specification changes could also affect the submitted code. Explaining this change requires further evidence about how students worked.

Higher lab correctness coincided with lower final-exam scores (Section~\ref{sec:results-rq2-cohort}), raising the question of whether success on the labs translates into stronger understanding. Differences in exam questions limit comparability across cohorts. Together with the decline in solution diversity, this contrast motivates closer attention to the relationship between task completion and learning. Novices who obtained complete solutions from a single prompt performed worse on subsequent unaided code-modification tasks~\cite{kazemitabaar2023use}, and randomized experiments on creative tasks found that assistance could improve immediate performance while reducing later independent performance and, in one condition, the diversity of ideas~\cite{kumar2025creativity}. These studies suggest examining how students use generated solutions and what they can subsequently do independently. For instructors, the practical question is whether a constrained task still elicits meaningful decisions that students must reason about, test, and explain.

The distinction between population-level patterns and individual attribution determines how the findings can be used. They support study of changes across cohorts in submitted code and motivate assignment review, but they provide no basis for questioning an individual student. Individual attribution would require held-out submissions with verified process histories, a reference bank and matching criteria specified before evaluation, and task-specific estimates of error rates~\cite{jegourel2024sieving,hoq2024detecting,pan2024detectors}.

\subsection{What Instructors Can Do Before Releasing an Assignment}

These findings support the use of generated-reference matching as a diagnostic before assignment release. Before students see an assignment, the instructor would generate solutions from the exact handout and starter code, evaluate every output with the instructor tests, and inspect convergence separately for each task. The resulting profile shows where an assignment channels solutions toward one form and where repeatedly generated failures expose ambiguity, giving the instructor a chance to revise the task before students encounter it.

A pilot reference bank is inexpensive relative to the generation run used in this study. At this study's average batch cost of about 5 US cents per attempt, 100 attempts would cost about US\$5. Our results do not establish that one hundred attempts are sufficient for every task, but the observed cost suggests that preliminary testing can be affordable. If validated outputs converge, an instructor can decide before release what additional evidence the assessment should elicit. Staged work, student-written tests, explanations of design choices, and short oral follow-ups can make reasoning and verification more visible~\cite{mahon2024guidelines,denny2024computing,kazemitabaar2023novice,nguyen2024misread}. Lab~10 also shows that low matching alone does not mean an assignment elicits richer reasoning, since different syntax inside one compact expression can lower the match rate. Regenerating the bank after a revision can then show whether the change altered model convergence under the same protocol.

Evidence of how students produced their code would strengthen future validation of generated-reference matching. AI-use disclosures could record which tool was used, for what purpose, which parts of the work it affected, and how the output was tested or revised~\cite{adnin2025undisclosed,ye2026disclosure}. Combining these disclosures with prompts, revisions, intermediate code, and student explanations would help researchers establish how submissions were produced and test whether final-code similarity reflects that process. Collecting this evidence would require appropriate consent and separation from decisions related to misconduct. The same records could also help instructors understand how students verify and revise their work.

%% file: sections/06-limitation.tex
\section{Limitations and Future Work}
\label{sec:threats}

This observational cohort comparison cannot validate individual AI-use detection, estimate LLM-use prevalence, or explain why the submitted code changed. No cohort has verified submission-level source labels or contemporaneous AI-use disclosures, and 2021 provides a historical reference whose individual source histories are unverified. Although the lead instructor, delivery mode, and core course materials were stable, other instructors, most teaching assistants, cohort composition, permitted support, and some checker or specification details varied. These differences prevent causal attribution of the observed shift.

The reference bank uses three 2026 frontier models under one protocol, so it measures retrospective resemblance to solutions generated by those models. It does not reconstruct the tools available to students in 2023 or 2025. Absolute similarity rates may change with future models, sampling, and preprocessing. MOSS reports surface overlap, and the student--LLM and LLM--LLM analyses use different docstring preprocessing. The exact-form estimates are limited to five purposively selected functions and remain sensitive to syntax. Neither hidden-test performance nor final-code similarity measures student understanding or effort, and the corpus contains final submissions from one Python CS1 course. Future work should evaluate preregistered reference banks on held-out code with verified process histories, replicate across courses, languages, and task structures, and collect prompts, revisions, intermediate code, and protected disclosures to connect final-code similarity with how students worked and what they learned.

%% file: sections/07-conclusion.tex
\section{Conclusion}
\label{sec:conclusion}

With LLM-generated code now readily available to students, instructors need to understand what matches between student submissions and generated reference solutions actually indicate. We examined 29,970 student submission files from ten CS1 labs across the 2021, 2023, and 2025 course offerings, comparing them with 90,000 reference attempts generated retrospectively by three frontier LLMs. The models usually produced correct solutions and often converged on the same implementations. Constrained tasks drew models and students toward canonical forms, whereas open-ended tasks retained more variation. Later cohorts more often matched the generated references, and four constrained functions appeared in fewer exact forms, including among full-pass submissions. These results document population-level change and provide evidence of students' solution convergence. Before assignment release, generated reference banks can reveal convergence or ambiguity and guide revisions that make reasoning, testing, and intermediate work visible. Determining how an individual submission was produced remains a separate question, one that requires student disclosure, verified process histories or other direct evidence.

%% file: sections/98-ack.tex
\begin{acks}
We would like to thank the Learning \& Education Advancement Fund (LEAF) from the Office of the Vice-Provost, Innovations in Undergraduate Education, University of Toronto, and the Natural Sciences and Engineering Research Council of Canada (NSERC) Discovery Grant (\#RGPIN-2024-04348). This work was supported by Research Council of Finland grants \#356114 and \#367787.

Generative AI was used responsibly in this research. During data analysis, we instructed Codex and Claude Code to assist in creating analysis and visualization scripts in Python. After we formed the initial interpretation of the data and authored a first draft of the manuscript, we employed an LLM to assist with copyediting and language formalization (to support the non-native English-speaking authors) and to cross-check for mismatches between data and claims throughout the paper. The authors take full responsibility for the final results and interpretations in the paper, having manually reviewed everything prior to the final submission.
\end{acks}

%% file: sections/99-appendix.tex
\section{Initial Generation Prompt}
\label{app:prompt-initial}
All LLMs received the same prompt printed below when generating lab responses:
\begin{lstlisting}[style=koliPrompt]
You are solving an introductory Python lab. Use the lab handout Markdown and starter file below. Return only the complete final contents of <STARTER_FILENAME>. Keep the original public names, signatures, imports, constants, and surrounding starter structure. Preserve existing docstrings and comments unless the handout or checker requires adding or correcting examples, doctests, or other documentation inside them. Only edit the parts that need to be edited, such as TODOs, pass statements, placeholders, incomplete function bodies, or docstrings that need required doctests. Do not include Markdown fences or explanations. Before returning, check the handout for required student tests, sanity checks, examples, or doctest counts, and satisfy those requirements in the returned file. When adding doctests, choose simple unambiguous examples and verify that each expected output matches the function specification and the implemented code.

Lab handout Markdown from <HANDOUT_FILENAME>:
~~~markdown
<LAB_HANDOUT_MARKDOWN>
~~~

Starter file:
```python
<STARTER_FILE_CONTENTS>
```
\end{lstlisting}